# Semiconductor Metasurfaces for Surface-enhanced Raman Scattering

*Haiyang Hu[1], Anil Kumar Pal[1], Alexander Berestennikov[1], Thomas Weber[1], Andrei Stefancu[1], Emiliano Cortés[1], Stefan A. Maier[1,2,3], and Andreas Tittl[1]\**.

1. Chair in Hybrid Nanosystems, Nanoinstitute Munich, Faculty of Physics, Ludwig-Maximilians-Universität München, Königinstraße 10, 80539 München, Germany.

2. School of Physics and Astronomy, Monash University Clayton Campus, Melbourne, Victoria 3800, Australia.

3. The Blackett Laboratory, Department of Physics, Imperial College London, London SW7 2AZ, United Kingdom.

\*E-mail: Andreas.Tittl@physik.uni-muenchen.de

**ABSTRACT**

Semiconductor-based surface-enhanced Raman spectroscopy (SERS) substrates, as a new frontier in the field of SERS, are hindered by their poor electromagnetic field confinement, and weak light-matter interaction. Metasurfaces, a class of 2D artificial materials based on the electromagnetic design of nanophotonic resonators, enable strong electromagnetic field enhancement and optical absorption engineering for a wide range of semiconductor materials. However, the engineering of semiconductor substrates into metasurfaces for improving SERS activity remains underexplored. Here, we develop an improved SERS metasurface platform that leverages the combination of titanium oxide ($TiO_2$) and the emerging physical concept of optical bound states in the continuum (BICs) to boost the Raman emission. Moreover, fine-tuning of BIC-assisted resonant absorption offers a pathway for maximizing the photoinduced charge transfer effect (PICT) in SERS. We achieve ultrahigh values of BIC-assisted electric field enhancement ($|E/E_0|^2 \approx 10^3$), challenging the preconception of weak electromagnetic (EM) field enhancement on semiconductor SERS substrates. Our BIC-assisted $TiO_2$ metasurface platform offers a new dimension in spectrally-tunable SERS with earth-abundant and bio-compatible semiconductor materials, beyond the traditional plasmonic ones.

**INTRODUCTION**

Surface-enhanced Raman spectroscopy (SERS) is a highly sensitive and non-destructive analytical technique that has garnered significant attention in the fields of biophysics, chemistry, and medical diagnostics due to its distinct vibrational fingerprints and molecular specificity.[1–5] The SERS substrate is at the core of SERS-based technologies, yielding large Raman enhancement factors through two independent mechanisms: electromagnetic (EM) and chemical mechanisms (CM).[2,6–8] So far, noble metals (Au, Ag, and Cu) played a dominant role in SERS applications due to their remarkable ability to quench fluorescence, and enhance the EM fields through plasmon resonances, achieving ultrasensitive and even single-molecule-level SERS detection.[9–11] However, there are some inevitable disadvantages for noble metal SERS substrates such as high cost, poor chemical and thermal stability, lack of uniformity and biocompatibility, low SERS signal reproducibility, and poor selective recognition ability for probe molecules.[9,12,13] Additionally, the strong photothermal effect may damage living cells, which can seriously restrict the utilization of noble-metal substrates in practical SERS platforms.[9] In contrast to noble metal substrates, semiconductor SERS substrates have emerged as an alternative due to their better chemical stability and biocompatibility[14]. Semiconductor materials possess distinct characteristics that endow them with exceptional capabilities, including the capacity to manipulate light at the nanoscale (like optical absorption engineering, light trapping by total internal reflection, etc.)[15,16], and their inherent adaptability allows for straightforward modification with a diverse range of functional groups ($-NH_2$, $-COO-$, $-PO_4^{3-}$, $-SH$, etc.)[2], which have significantly broadened SERS applications across various fields.[4,9,13,17] In particular, semiconductor SERS substrates benefiting from photoinduced charge transfer (PICT),[9,13,18] have also exhibited high analytical sensitivity. These substrates have demonstrated great potential for detecting target molecules in complex environments and monitoring interfacial chemical reactions, making them valuable tools in the fields of biophysics, chemistry, and medical diagnostics research.[19]

However, compared to noble metals, semiconductor substrates generally have a lower limit of detection, which can be attributed to weaker EM field enhancement and light-matter interactions.[12,20–22] Therefore, it is crucial to explore strategies to improve the SERS performance of semiconductor substrates.[23] To this end, several methods have been reported to enhance the EM field and light-matter interaction. Common approaches based on Mie resonances are to synthesize semiconductor nanospheres to trap light inside,[24] or assemble homogeneous spheres to form two-dimensional arrays or three-dimensional colloidal crystals.[25] Alternatively, semiconductor doping has also been applied to generate surface plasmon resonance (SPR), which inevitably introduces enormous optical loss.[2,26] However, these methods have not fully harnessed one of the most distinctive advantages of semiconductors: their optical engineering capabilities.[2,27] In this regard, semiconductors can act as versatile building blocks for developing 2D monolithic metasurfaces, offering exciting possibilities for managing and manipulating light at the interfaces. Such advancements hold the potential to pave the way for breakthroughs in next-generation SERS platforms.[2,28,29]

Metasurfaces constructed from two-dimensional subwavelength arrays of semiconductor nanostructures have shown tremendous potential for enhancing light-matter interaction at the nanoscale.[30,31] In particular, metasurfaces underpinned by the physics of bound states in the continuum (BIC) have seen a surging interest due to their strong light confinement and remarkable enhancement of electromagnetic fields,[32,33] stimulating applications across diverse fields, including nanoscale lasing,[34,35] biomolecular sensing,[36,37] and nonlinear photonics.[38,39] Hence, BIC-assisted metasurfaces serve as an ideal toolkit for light manipulation in semiconductor SERS substrates (Figure 1a), which, to the best of our knowledge, have not yet been realized.

Here, we demonstrate BIC-driven semiconductor metasurfaces as a new platform for high-sensitivity SERS. Through precise nanophotonic engineering of quasi-BIC resonances in the TiO$_2$ BIC system, we successfully enhance both the PICT effect (chemical mechanism) and the EM field confinement (physical mechanism) of SERS. Significantly, the integration of metasurface nano-engineering techniques with BIC physics offers a versatile approach applicable to a wide range of semiconductor SERS substrate materials (with low intrinsic losses), which can adapt well to diverse molecules like methylene blue (MB), Rhodamine 6G, and adenine.[9,13,40] This synergistic combination leverages the benefits of nanoscale light manipulation (metasurface) and the remarkable chemical effects empowered by photoinduced charge transfer (semiconductor), thereby optimizing the SERS performance of these substrates.

**RESULTS AND DISCUSSION**

Conceptually, a BIC is a localized state existing in a continuum of radiative modes.[32,41] This phenomenon originally appeared in quantum mechanics and has later been applied to many other areas of physics where resonant behavior is prevalent.[42] A true BIC, with an infinite value of quality factor (Q factor, defined as the resonance position divided by the line width), can be explained by vanishing coupling constants with all radiation channels.

One way of making BICs usable in practical nanophotonic systems is to design symmetry-protected metasurfaces, where the coupling constants to the radiation continuum are tailored using structural asymmetry within the metasurface unit cell, leading to the formation of quasi-BIC modes accessible from the far field.[32] The narrow quasi-BIC could be a drawback from the perspective of scaling the EM enhancement of the Stokes-shifted field, however, it offers several advantages for the SERS application. A key advantage of symmetry-protected BICs is their ability to accurately control the resonance position (Figure 1b). This attribute is particularly crucial for semiconductor SERS platforms, since matching the quasi-BIC resonance to the laser wavelength can significantly enhance the SERS sensitivity.

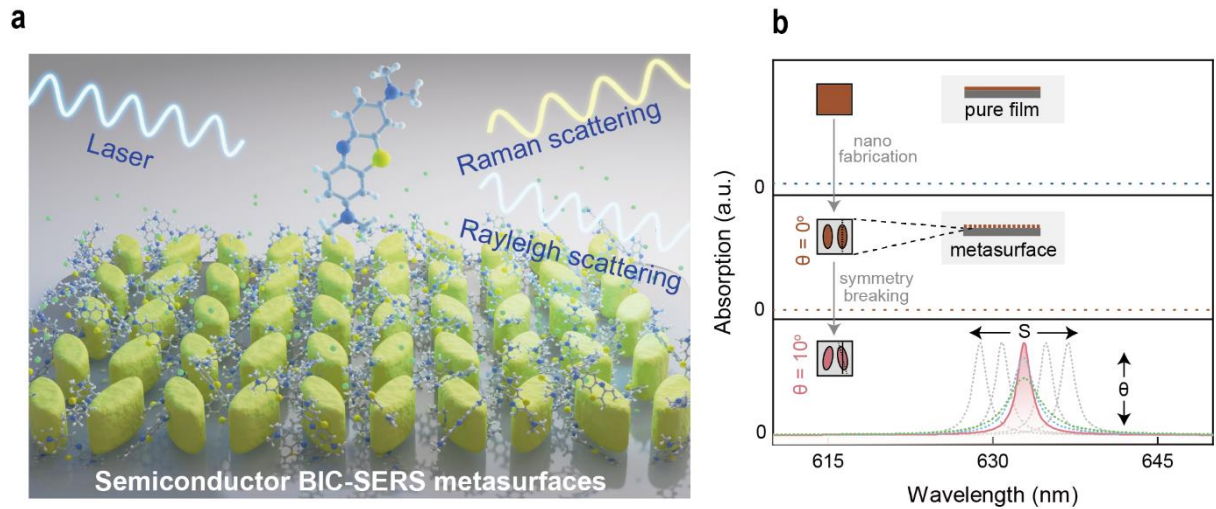

**Figure 1. BIC-driven TiO$_2$ semiconductor SERS metasurface.** (a) Schematic of the TiO$_2$-based BIC metasurface platform for SERS applications. (b) Illustration of BIC-assisted absorption enhancement and spectral tunability. The simulated absorbance spectra for the pure TiO$_2$ film (top), BIC-assisted TiO$_2$ metasurfaces with the symmetrical geometry (middle), and asymmetrical geometry ($\theta = 10°$) as an example (bottom). The absorbance of the metasurface can be enhanced with the excitation of quasi-BIC resonances in asymmetrical geometry. This can be further spectrally tailored through the $S$ (scaling factor), and the amplitudes can additionally be tuned by adjusting the degree of asymmetry factor ($\theta$).

**Numerical Investigation of TiO$_2$ BIC metasurfaces.** We implemented metasurfaces with two different geometries (two ellipses, and hole-in-disk) for our BIC metasurfaces designs (Figure 2a, b). As introduced previously, for semiconductor SERS applications, both enhanced absorption of the incident light source leading to the substrate-molecule PICT effect, and the inherent strong electric field enhancement of the nanomaterial are vital to the amplification of the Raman signal. Hence, in our simulation, we evaluate the absorption and electric field enhancement for the metasurfaces featuring with these two different geometries.

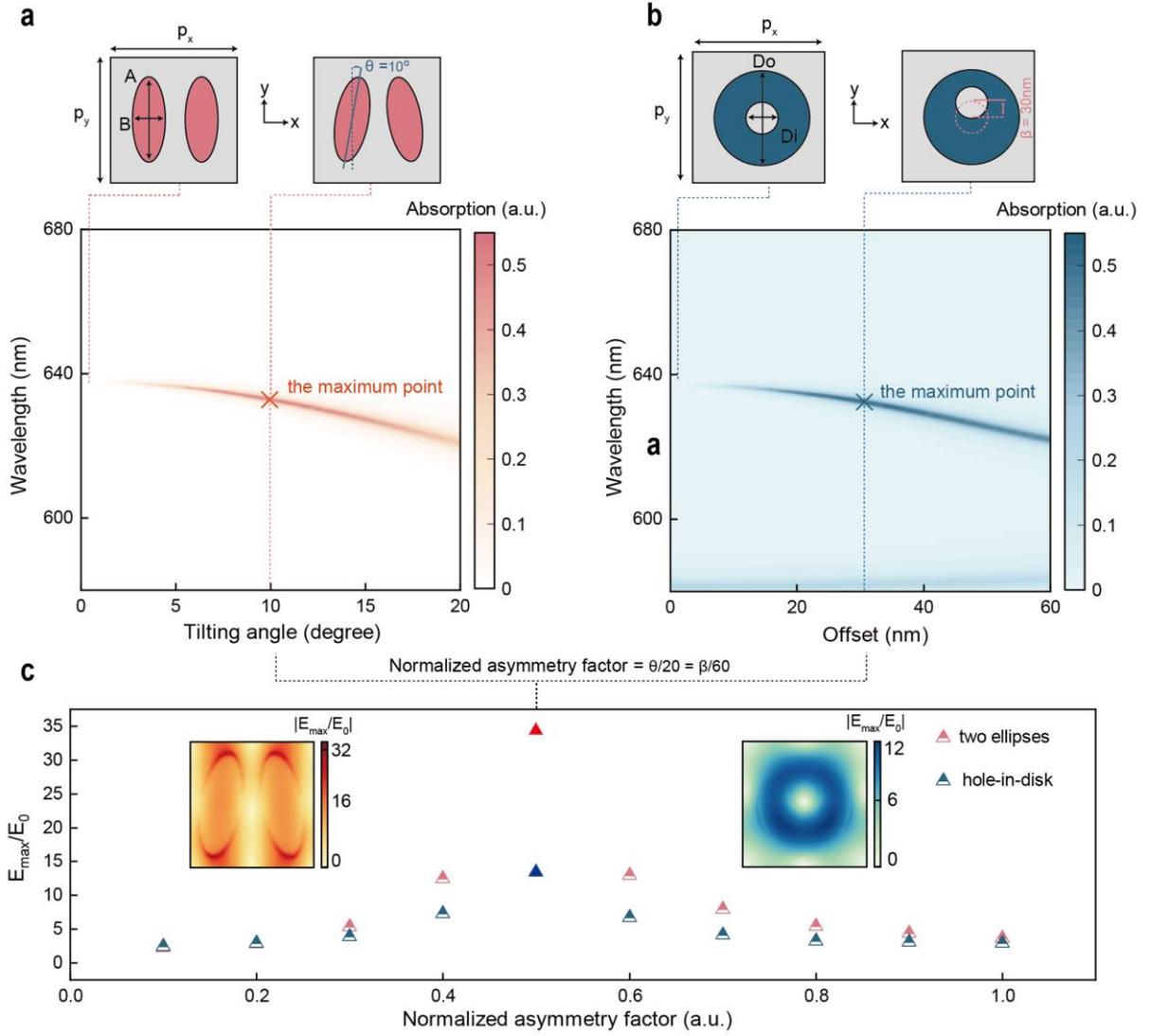

**Figure 2. Numerical design of semiconductor SERS metasurfaces and their far/near field properties.** (a) Top: Sketch of a BIC unit cell with the two ellipses geometry (A, B, $P_x$, and $P_y$ will scale linearly according to scaling factor *S*). The geometrical parameters of the unit cell are A = 302nm, B = 115nm, $P_x$ = $P_y$ = 366 nm (*S* = 1). The symmetry can be broken by tuning the tilting angle (*θ*). Bottom: Color-coded simulated absorbance maps of $TiO_2$ metasurfaces as a function of tilting angle *θ* and wavelength. (b) Top: Sketch of the unit cell with hole-in-disk geometry (Di, Do, $P_x$, and $P_y$ will scale linearly according to scaling factor *S*). The geometrical parameters of the unit cell are Di = 90 nm, Do = 263 nm, $P_x$ = $P_y$ = 366 nm (*S* = 1). Here we can shift the hole along y axis (*β*) to break the symmetry. Bottom: Color-coded simulated absorbance maps of metasurfaces as a function of offset value (*β*) and wavelength. (c) The simulated electric field enhancement of metasurfaces with different geometries (two ellipses, and hole-in-disk) under different asymmetry factors. For ellipses and hole-in-disk geometries, we normalize the asymmetry parameter $\theta$, and $\beta$ through, normalize asymmetry factor = $\theta$ / 20 = $\beta$ / 60. Here we set the tuning range of $\theta$ from 0 to 20 degrees, and $\beta$ from 0 to 60 nm, which are consistent with the above simulated absorbance maps. Inset: The electric near-field at the resonance frequency for unit cells (two ellipses, and hole-in-disk) of $TiO_2$ metasurfaces, corresponding to maximum absorption, and field enhancement.

As mentioned above, a true BIC is a mathematical concept characterized by infinite quality factor and vanishing spectral line width. In practice, true BICs can be converted to quasi-BICs with finite resonance width by breaking the in-plane inversion symmetry within the unit cell, allowing the resonant mode to be excited from the far field. For the unit cell with two ellipses, the symmetry breaking can be achieved by tilting the ellipses at an angle $\theta$ with respect to each other. In the hole-in-disk geometry, the symmetry is broken by shifting the hole from the center along the y-axis by a specific value ($\beta$).

Through the quasi-BIC resonance, we can achieve a significant enhancement in absorption after breaking the $C_2^z$ symmetry in the metasurfaces. The absorbance spectrum are derived from the simulated reflectance ($R$) and transmittance ($T$) spectra ($A = 1 - R - T$). Here, we emphasize one of the most important advantages of BIC metasurfaces in SERS applications, namely that the light confinement at the interface (characterized by absorbance) can be maximized through tuning the asymmetry parameter. Specifically, numerical simulations reveal absorbance maps of $TiO_2$ metasurfaces with different geometries (two ellipses, and hole-in-disk) as a function of the asymmetry parameters ($\theta$, and $\beta$, respectively). As can be seen in the simulated absorption maps (Figure 2a, b), on both metasurfaces, the absorbance enhancement of quasi-BIC resonances is a function of asymmetry parameter, reaching maximum magnitudes at specific values ($\theta$ = 10 degrees, and $\beta$ = 30 nm, respectively), as determined by the critical coupling condition explained by temporal coupled-mode theory. In brief, the BIC-enabled semiconductor SERS metasurface platform can be described as a single-mode cavity with two mirror-symmetric ports, where the interaction between the far-field and the cavity mode at each port is determined by the coupling rate $\kappa$, which can be derived from the radiative decay rate $\gamma_{rad}$. Additionally, an intrinsic decay rate $\gamma_{int}$ accounts for the energy loss through material absorption, and the far-field absorbance can be calculated according to:

$$A = \frac{2\gamma_{int}\gamma_{rad}}{(\gamma_{int}+\gamma_{rad})^2} \qquad \text{Equation (1)}$$

which reaches its maximum value when the intrinsic decay rate $\gamma_{int}$ is equal to the radiative decay rate $\gamma_{rad}$, referring to the critical coupling condition.[43]

Although both geometries (two ellipses, and hole-in-disk) can achieve the maximum absorbance magnitude (A = 0.5) under the critical coupling condition, they present different near-field EM enhancements, which is also an important aspect for SERS.[44] It can be clearly seen that the unit cell with two ellipses geometry exhibited stronger electric field enhancement ($|E/E_0| = 32$) than hole-in-disk geometry ($|E/E_0| = 12$) even though both of them are under the critical coupling condition (Figure 2c). It's worth noting that semiconductor SERS substrates typically rely on the chemical mechanism (PICT) to amplify the Raman signal, but often suffer from extremely weak EM enhancement.[12,20,21] Here, our semiconductor metasurfaces, empowered by the BIC resonances, allow for strong electric field enhancement. Conducting the SERS experiment on these metasurfaces of different geometries could deepen our understanding of how the BIC metasurface geometry affects SERS performance.

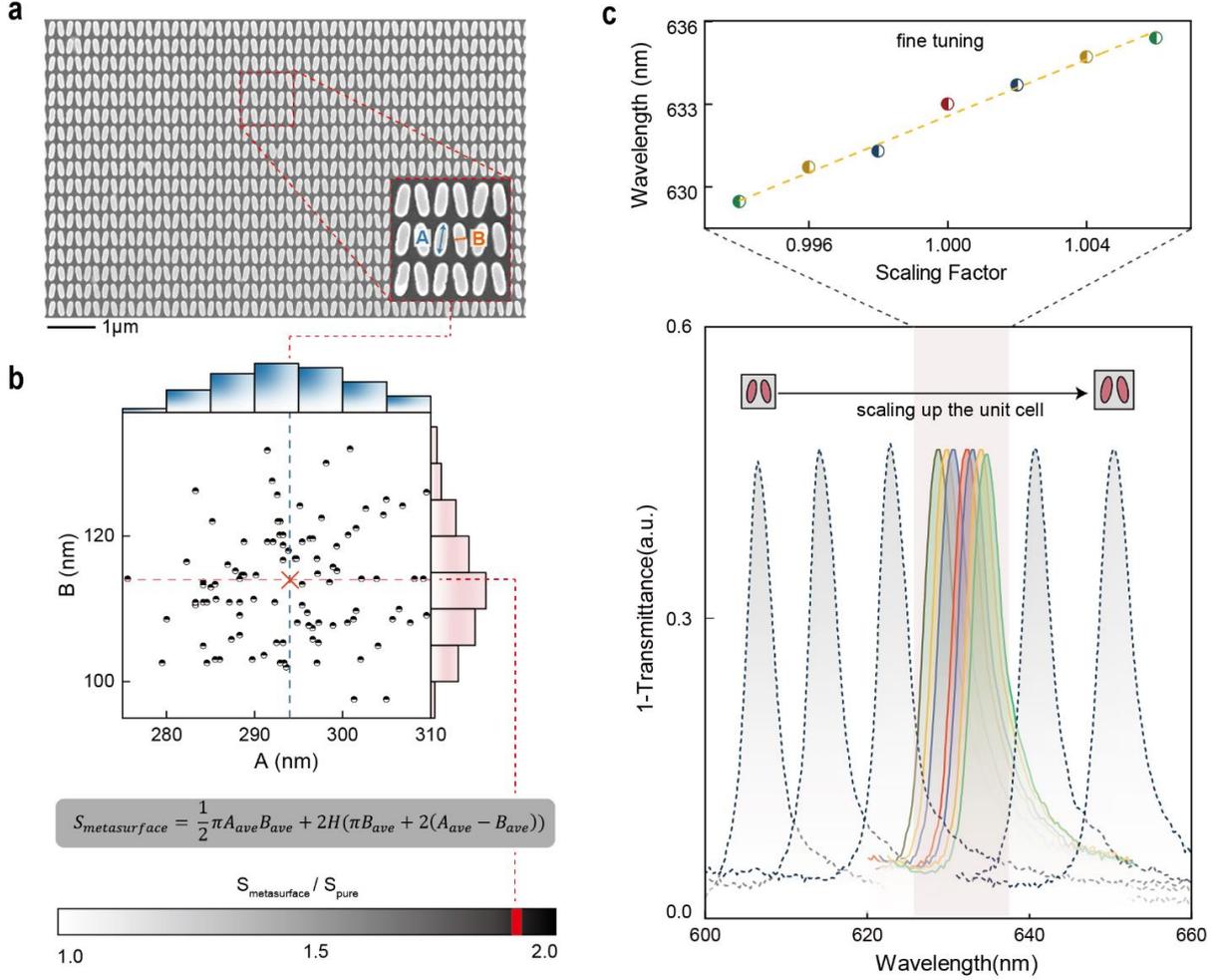

**Figure 3. Experimental metasurface fabrication and spectral tuning.** (a) SEM images of experimental metasurfaces with two ellipse geometry (scaling factor $S = 1$). (b) Statistical measurements of the length of long and short axes (A and B) of the ellipses in different unit cells (100 in counting), where average values for A and B are 294 and 114 nm, respectively. The surface area of a single unit cell ($S_{metasurface}$) is calculated based on these values according to the proximate formula shown in the figure, which is compared with the surface area of the unstructured pure film ($S_{pure}$) with the same size as the periodic of the single unit cell (366 × 366nm), resulting in a factor of 1.9 ($S_{metasurface}/S_{pure}$). (c) Measured transmittance spectra for metasurfaces with different scaling factors (0.994 to 1.006 for fine tuning).

**Experimental fabrication of TiO$_2$-Based metasurfaces.** TiO$_2$ thin films with a thickness of 140 nm were fabricated by sputter deposition. To fabricate TiO$_2$-based metasurfaces, we employed top-down nanofabrication through high-resolution electron beam lithography and anisotropic reactive ion etching (see the "Method" section). We experimentally implemented the metasurfaces designs composed of two ellipses (Figure 3a) and hole-in-disk geometries (Figure S1). Here, we took two ellipses metasurfaces as an example to confirm their fabrication quality using electron microscopy (Figure 3a). We performed statistical measurements on the long and short axes (A and B) of 100 unit cells, obtaining average values of ±294 and ±144 nm, respectively (Figure 3b). We increased the effective surface area of semiconductor metasurfaces by approximately 2 times compared to the non-structured pure $TiO_2$ film, which provides a greater active contact surface area for analyte molecules to form surface complexes and generate stronger Raman signal.[40]

The presence of quasi-BIC resonances has been through white light transmittance measurements (see the "Method" section). We experimentally demonstrate broad spectral tunability of the semiconductor BIC metasurfaces by scaling the size of the unit cell (*S*), which enabled us to control the position of the quasi-BIC resonance throughout the whole visible spectrum (Figure 3b), and cover the widely used visible Raman laser sources (532, 633, 785 nm). Fine-tuning the *S* factor around the target laser wavelength (633nm here) ensured that we could precisely excite the quasi-BIC resonance. We applied the temporal coupled-mode theory (TCMT) model to fit the experimental transmittance spectrum,[45] which characterizes the spectral mode energy density and is directly related to the enhancement of light-matter interaction[46].

Matching the excitation wavelength to the semiconductor substrate absorption has been a long-standing challenge for SERS semiconductor substrates,[3] which has now been experimentally achieved in our semiconductor BIC metasurfaces. Additionally, when the laser wavelength matches the molecular resonance, the Raman signal can be amplified by several orders of magnitude due to the resonant Raman effect (as discussed in the following section).[44]

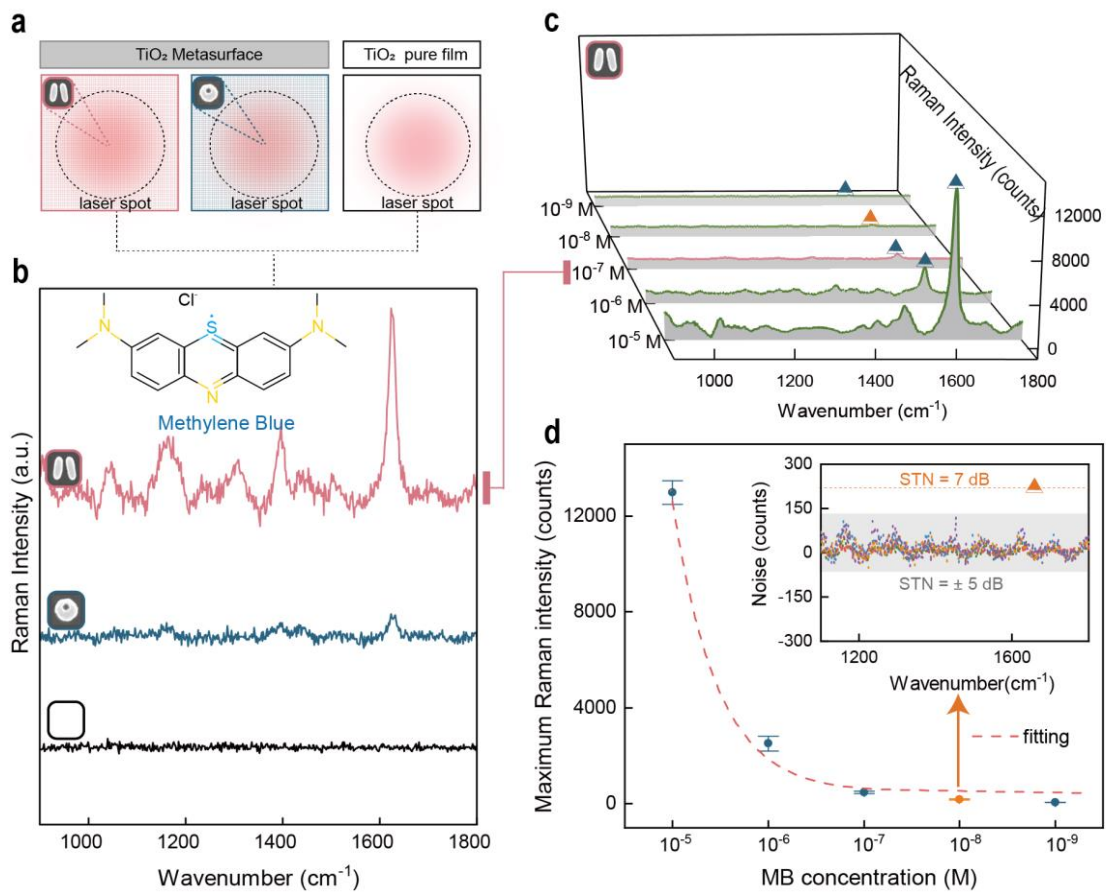

**Figure 4. Raman spectra of methylene blue on the BIC-assisted semiconductor SERS metasurface.** (a) Schematic layout of the laser illuminating on the $TiO_2$ BIC metasurfaces (two ellipses, and hole-in-disk geometries), and the pure $TiO_2$ film, where BIC metasurfaces pattern ($32 \times 32$ μm$^2$) can be successfully excited by a laser (14 μm in diameter). (b) Raman spectra for MB were measured on the BIC metasurfaces with different geometries (two ellipses, and a hole-in-disk), and on the pure $TiO_2$ film, where the concentration of MB solution is $10^{-7}$ M, and the excitation laser wavelength is 633nm. (c) The Raman spectra for MB at different concentrations ($10^{-5}$ to $10^{-9}$ M) on the BIC-assisted $TiO_2$ semiconductor SERS substrate with two ellipses geometry. (d) The maximum Raman signal of MB (at 1624 cm$^{-1}$) at different concentrations measured on the metasurfaces with two ellipses geometry (data extracted from Figure 4c), and the analysis of the signal-to-noise ratio (inset). The noise signal background is gathered from the unstructured $TiO_2$ film under the same settings.

**BIC-assisted semiconductor SERS metasurface.** We utilized MB as a test molecule for SERS investigations together with the 633 nm laser as the excitation source (see Methods). Taking advantage of the high spectral tunability of our semiconductor SERS substrate, we engineered the resonance position of metasurfaces with different meta-units (two ellipses, and a hole-in-disk) to match the pump wavelength (633nm) by finely tuning the *S* factor (discussed in Figure 3). The target BIC metasurfaces pattern with the size of $33 \times 33$ μm$^2$ is illuminated by the excitation laser with a spot size around 14 μm in diameter passing through a 20 X water-immersed objective, which ensured the successful excitation of quasi-BIC resonance (Figure 4a). It is well known that in practical SERS applications, the molecular fluorescence signal can be recorded during the measurement. This is particularly important in semiconductor SERS measurements, unlike metals, there is no quenching of the fluorescence in this case.[23] Hence, in this research, we employed an automated algorithm for fluorescence removal based on modified multi-polynomial fitting (see Methods, and Figure S2a,b).[47]

To better understand the SERS application of the BIC-assisted semiconductor TiO$_2$ metasurfaces, we compared the SERS enhancement on metasurfaces with different geometries (two ellipses, and hole-in-disk). Specifically, under the same MB concentration ($10^{-7}$ M), the semiconductor BIC metasurface with two ellipses exhibited a SERS signal approximately 7 times higher than that of hole-in-disk geometry, which is attributed to the higher EM field enhancement associated with the two ellipses geometry (discussed in next section) in the vicinity of BIC metasurfaces. The outstanding SERS enhancement by the BIC metasurfaces was clearly observed when we compared it with the non-patterned pure TiO$_2$ film where no appreciable Raman scattering was detected, thereby proving the fundamental advantages offered by the BIC metasurfaces engineering technique for semiconductor SERS substrates. Additionally, to confirm that the SERS enhancement is indeed induced by the BIC resonance, we deactivate the resonance by rotating BIC metasurfaces at 90º thus mismatching the polarization of the BIC with the laser, which consequently resulted in no SERS enhancement (Figure S2c), despite having the same surface (i.e. equal amount of molecules interacting with the platform).

The detection sensitivity of the semiconductor BIC metasurfaces with two ellipses geometry for SERS was further investigated by performing Raman measurements at different MB concentrations ($10^{-5}$ to $10^{-9}$ M). It can be clearly seen that the Raman signal intensity decreased sharply with decreasing MB concentration (Figure 4c). The detection sensitivity for MB on our BIC-assisted semiconductor SERS metasurfaces was $10^{-8}$ M, which was evaluated from the signal-to-noise ratio, in accordance with IUPAC recommendations.[48] We estimate the detection sensitivity of our SERS metasurfaces by considering both the processing error induced by mathematically removing the autofluorescence background signals (see Methods) and the noise level associated with our experiments, which includes detector noise, source stability, and optical alignment.[37] Specifically, we captured a total of $n = 5$ reference Raman measurements, using BIC metasurfaces that mismatched the excitation laser (633nm) and proceeded with the same mathematical background noise removal treatment. Based on the obtained Raman spectra, we calculated the standard deviation for each reference Raman spectrum and obtained the average value for the amplitude of the noise background (see Methods). The signal-to-noise ratio (STN, in dB) was calculated according to the following formula (Equation 2), where $A_{signal}$ is the amplitude of the specified Raman signal (1624 cm$^{-1}$) mode, and $A_{noise}$ is the amplitude of background noise, which can be obtained as described earlier.

$$STN = 10 \times log_{10}(A_{signal}/A_{noise}) \qquad \text{Equation (2)}$$

The calculated STN of the Raman mode (1624 cm$^{-1}$) with maximum amplitude is approximately 7dB (Figure 4d), which exceeds the common threshold for minimum acceptable SNR (3 dB)[49].

In summary, we have experimentally demonstrated significant SERS enhancement of MB on our BIC-assisted semiconductor metasurfaces, achieving detection sensitivity of 10$^{-8}$ M which is higher than the reported TiO$_2$ semiconductor SERS substrates for MB sensing (10$^{-6}$ M).[20,21] To deepen our understanding of mechanisms underlying the SERS enhancement achieved through BIC-assisted semiconductors metasurfaces, we will discuss the chemical and EM enhancement mechanisms in the following section.

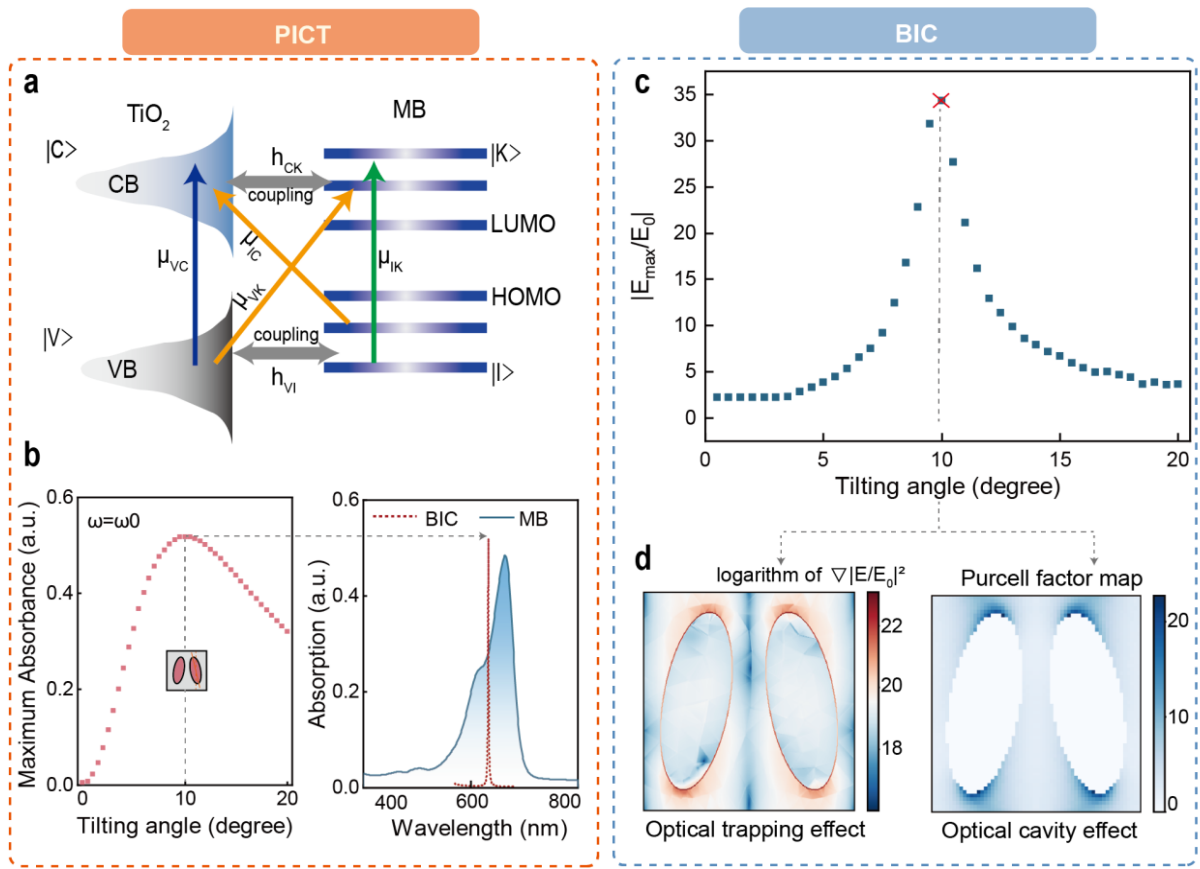

**Figure 5. Hybrid effect of BIC-assisted semiconductor SERS metasurface.** (a) Schematic PICT transition in the semiconductor-molecule system. (b) Left: Simulations of absorbance at the resonance ($\omega=\omega_0$) for the TiO$_2$ BIC metasurfaces (two ellipses geometry) at different tilting angles ($\theta$ = 1-20°). Right: Enhanced absorption of TiO$_2$ BIC metasurfaces ($\theta$ = 10°), aligned with the absorption of MB (10$^{-3}$ M). (c) Simulations of the electric field enhancement in TiO$_2$ BIC metasurfaces at various tilting angles ($\theta$ = 1-20°), where the maximum field enhancement is achieved at $\theta$ = 10°, satisfying the critical coupling condition (radiation loss equals intrinsic loss). (d) Left: The logarithm of the gradient of the squared electric field enhancement, based on simulation results, which can be used to estimate the optical trapping effect on the MB molecular. Right: Map of the Purcell factor values for a single unit cell to evaluate the enhanced light-matter interaction in the BIC nanocavity.

**Hybrid effects on BIC-assisted semiconductor SERS metasurface.** It is well known that enhancing the photoinduced charge transfer (PICT) effect between the semiconductor and molecule is a general strategy for achieving a high SERS enhancement on semiconductor SERS substrates.[50] Specifically, when the molecules are bound to the semiconductor surface by weak covalent bonds, the charge can be transferred across the interface between the semiconductor and molecule, which can generate strong Raman enhancement if the excitation frequency resonates with the PICT transition.[2,18,51,52] Specifically, when the energy levels of the semiconductor are coupled with the absorbed molecules, the charge can be transferred from the valence band states (|V>) of the semiconductor to the molecular excited states (|K>) via the transition moment, $\mu_{VK}$, or from the molecular ground state (|I>) to the conduction band states (|C>) of the semiconductor via the transition moment, $\mu_{IC}$, which can be further enhanced through borrowed intensity either from exciton transition $\mu_{VC}$ or molecular transitions $\mu_{IK}$, through the Herzberg-Teller coupling term ($h_{CK}$ or $h_{VI}$), as shown in Figure 5a.[18] The highly efficient PICT process between the TiO$_2$-molecules has been confirmed by many reported works.[18,20]

As discussed above, matching the excitation laser with the molecular resonance is pivotal for achieving high PICT rates, and consequently high SERS enhancement factors, typically by several orders of magnitude (3-4 orders).[2] This enhancement can be further amplified by SERS substrates, if their absorption of laser source is maximized. Due to the diversity of molecular analytes, conventional semiconductor SERS substrates face challenges in achieving a maximized absorption of the excitation laser source which matches the molecular resonance. This requirement can now be met through utilizing the metasurface engineering technique, assisted by BIC physics. Specifically, the absorption of $TiO_2$ metasurface (two ellipses geometry) can be optimized to its maximum value by precisely tailoring the radiation loss via tuning the asymmetry factor ($\theta$) to match intrinsic losses, thus satisfying the critical coupling condition, which has been systematically investigated in our previous work.[53] Here, we obtained the maximum absorption on the $TiO_2$ metasurface with $\theta = 10^o$ (Figure 5b). Taking advantage of the remarkable spectral tunability of $TiO_2$ BIC metasurfaces, we were able to precisely tune the maximum absorption of the excitation wavelength (633nm) to the molecular resonance of MB, by adjusting the scaling factor of the unit cell (as discussed in Figure 4c). Once successfully achieving the maximum absorption enhancement of $TiO_2$ metasurface at the molecular resonance position, the charge-transfer resonance and molecular resonance can couple to each other by the Herzberg-Teller coupling constant[2], which can further boost the PICT effect and provide stronger SERS activity of our BIC-assisted semiconductor SERS metasurface.

Even though the chemical mechanism (PICT) plays an important role in semiconductor SERS technologies, here, we also emphasize the importance of the physical mechanism brought by BIC metasurfaces to the field of semiconductor SERS substrates.

As the SERS signal is proportional to the enhancement of the EM field,[54] achieving high EM field enhancements has been a long-standing challenge in the semiconductor SERS community as it is highly reduced in these systems the possibility of exciting localized surface plasmon, compared to their metal counterparts. However, we can achieve strong EM field enhancement by engineering the semiconductor TiO$_2$ substrate into a BIC metasurface, thereby benefiting from strong light localization provided by the BIC resonances. As shown here, based on our TiO$_2$ substrate, the maximum EM field enhancement can be achieved with $\theta = 10º$ (Figure 5c), where |E| represents the absolute magnitude of the local electric field with a metasurface, and |E$_0$| represents incident field without structure, for clarity. The highest electric field enhancement (|E|/|E$_0$|) here is around 35 (larger than hole-in-disk geometry) at the resonance frequency. Additionally, the large gradient of the squared electrical field $F_{grad} \propto \nabla |\vec{E}(r)|^2$ at the hot spots region (Figure 5d) suggests the possible molecule trapping effect.[55] The contribution of enhanced EM field to the amplification of radiative spontaneous emission can be explained by the Purcell effect (Equation 3),[56] the influence of which has been demonstrated in SERS studies:[46]

$$F_p = \frac{3}{4\pi^2}\left(\frac{\lambda_c}{n_c}\right)^3 \left(\frac{Q}{V_{eff}}\right) \qquad \text{Equation (3)}$$

Figure 5d shows the calculation of the Purcell factor map in a unit cell at the resonant frequency (details in Methods). Comparing the values of the Purcell factor, one can clearly see that the Purcell factor around the vertices of the ellipses is much higher (F$_p$ ~ 20).

**CONCLUSION**

In this study, we have developed TiO$_2$-based semiconductor SERS metasurfaces, and demonstrated that the synergistic integration of semiconductor materials and BIC-resonant photonic metasurfaces offers tremendous potential for advanced sensing and detection applications. We challenge the preconception that the conventional semiconductor SERS substrates suffer from inefficient EM field enhancement, by showing high electric field enhancement ($|E/E_0|^2 \approx 10^3$) on our BIC platform. Additionally, due to the possibility of fine-tuning the BIC resonance across a wide range, it is easy to match it with the absorption wavelength of the analyzed molecule resulting in the enhanced PICT effect. Based on our BIC-assisted semiconductor SERS metasurface, we demonstrate a detection efficiency for the molecule MB at $10^{-8}$ M, two-fold higher than the one reported for $TiO_2$ semiconductor SERS substrates ($10^{-6}$ M). [20,21] Finally, our BIC-assisted semiconductor SERS metasurface platform circumvents the long-standing drawbacks of many semiconductor SERS substrates, such as the poor spectral tunability of light absorption, weak electromagnetic field enhancement, and lack of strong light-matter interaction, and shows that improved SERS activity can be extended to multiple sensing systems with different semiconductor substrates and analytes.

**METHODS**

**Numerical simulation:** Simulations were performed using the finite-element frequency-domain Maxwell solver included in CST Studio Suite 2021, and Lumerical FDTD software taking into account the experimentally measured optical constants of $TiO_2$. We utilized the default value (n = 1.5) implemented in CST Studio Suite for the $SiO_2$ substrate. The optical far-field absorbance in an aqueous environment was calculated via $A = 1 - T - R$, where transmittance ($T$) and reflectance ($R$) spectra were simulated under linearly polarized normally incident illumination. The E-field monitor with the corresponding eigenfrequency was added for the electric near-field simulation. The Purcell factor was calculated using the COMSOL Multiphysics package. A dipole source located along the y axis was moved along the central unit cell of the finite (15 × 15 unit cells) BIC metasurface. The Purcell factor was calculated as the ratio of the integral power flux through the sphere surrounding the dipole in the presence of a metasurface and in a vacuum.

**Nanofabrication of metasurface:** Sputter deposition (Angstrom) was applied to produce TiO$_2$ (140 nm) films. Before the electron-beam lithography step, the sample was spin-coated with a layer of photoresist (PMMA 950K A4) followed by spin-coating a conducting layer (ESpacer 300Z). The unit cells with different geometries (two ellipses, and hole-in-disk) were patterned with electron-beam lithography (Reith Eline Plus) with an acceleration voltage of 20 kV and a 15 μm aperture. The samples were transferred to a 3:1 MIBK: IPA solution for developing for 135s, followed by deposition of a 50 nm chrome layer as the hard mask. Lift-off was conducted in Microposit Remover 1165 overnight at 80 °C, followed by reactive ion dry etching in an RCP-RIE system with a SF$_6$/Ar plasma for 135s. The chrome hard mask was removed through wet etching with chromium etchant (Sigma-Aldrich).

**Optical measurements:** Transmittance measurements of the fabricated metasurface samples were carried out with a WiTec optical microscope comprising a water immersion objective (20X, NA = 0.5, Zeiss, Germany), where the metasurfaces were immersed in methylene blue solution ($10^{-7}$ M, 150 μL) illuminated by a Thorlabs OSL2 white light source with linear polarization.

**SERS measurement:** Raman measurements were carried out with the WiTec optical microscope equipped with three CW lasers 532, 633, and 785 nm. The TiO$_2$ BIC metasurfaces were immersed in 150 μL of methylene blue (Sigma-Aldrich) water solution ($10^{-7}$ M) illuminated with 633 nm laser (spot size of 14 μm in diameter) through a water immersion objective (20X, NA = 0.5, Zeiss, Germany) on the corresponding metasurfaces pattern (33×33 μm$^2$) with matching resonance wavelength. For spectra acquisition, the incident power on the sample was set to 1 mW with a dwell time of 50 s. The Raman signal was computationally derived from the fluorescence background through literature reported subtraction algorithm, where modified multi-polynomial fitting was applied, which suppresses the undesirable artificial peaks that might occur in polynomial fitting though taking the effects of noise level and peak contribution into account.

**AUTHOR INFORMATION**

Corresponding Author

*E-mail: Andreas.Tittl@physik.uni-muenchen.de

**ACKNOWLEDGEMENTS**

This work was funded by the Deutsche Forschungsgemeinschaft (DFG, German Research Foundation) under grant numbers EXC 2089/1 – 390776260 (Germany's Excellence Strategy) and TI 1063/1 (Emmy Noether Program), the Bavarian program Solar Energies Go Hybrid (SolTech), and the Center for NanoScience (CeNS). A.S. acknowledges the Alexander von Humboldt Foundation for a postdoctoral fellowship. S.A. Maier additionally acknowledges the EPSRC (EP/W017075/1), the Australian Research Council, and the Lee-Lucas Chair in Physics. Funded by the European Union (ERCs, CATALIGHT 802989 and METANEXT 101078018). Views and opinions expressed are however those of the author(s) only and do not necessarily reflect those of the European Union or the European Research Council Executive Agency. Neither the European Union nor the granting authority can be held responsible for them.

**AUTHOR CONTRIBUTIONS**

H.H. conceived, designed, and implemented the experiments. A.K.P. optimized the experiment design and analyzed the data. A.B. performed the simulation of Purcell factor. T.W. did the theoretical analysis of TCMT. A.S. and E.C. helped with experiment design and reviewing. S.A.M and A.T. conducted project administration and supervision. All authors contributed extensively to the work presented in this paper.

**COMPETING FINANCIAL INTERESTS**

The authors declare no competing financial interest.

**DATA AVAILABILITY**

All relevant data are available from the corresponding author on request.

**REFERENCES**

1. Li, J. F. *et al.* Shell-isolated nanoparticle-enhanced Raman spectroscopy. *Nature* **464**, 392–395 (2010).

2. Alessandri, I. & Lombardi, J. R. Enhanced Raman scattering with dielectrics. *Chem Rev* **116**, 14921–14981 (2016).

3. Sun, F. *et al.* Hierarchical zwitterionic modification of a SERS substrate enables real-time drug monitoring in blood plasma. *Nat Commun* **7**, 13437 (2016).

4. Haldavnekar, R., Venkatakrishnan, K. & Tan, B. Non plasmonic semiconductor quantum SERS probe as a pathway for in vitro cancer detection. *Nat Commun* **9**, 3065 (2018).

5. Arabi, M. *et al.* Chiral molecular imprinting-based SERS detection strategy for absolute enantiomeric discrimination. *Nat Commun* **13**, 5757 (2022).

6. Otto, A. The 'chemical' (electronic) contribution to surface-enhanced Raman scattering. *J Raman Spectrosc* **36**, 497–509 (2005).

7. Le Ru, E. C., Blackie, E., Meyer, M. & Etchegoint, P. G. Surface enhanced Raman scattering enhancement factors: A comprehensive study. *J Phys Chem C* **111**, 13794–13803 (2007).

8.	Le Ru, E. C., Etchegoin, P. G. & Meyer, M. Enhancement factor distribution around a single surface-enhanced Raman scattering hot spot and its relation to single molecule detection. *J. Chem Phys* **125**, (2006).

9.	Zheng, Z. *et al.* Semiconductor SERS enhancement enabled by oxygen incorporation. *Nat Commun* **8**, 1993 (2017).

10.	Lin, B. W. *et al.* Aluminum-black silicon plasmonic nano-eggs structure for deep-UV surface-enhanced resonance Raman spectroscopy. *Appl Phys Lett* **120**, (2022).

11.	Wang, Z., Ho, Y. L., Cao, T., Yatsui, T. & Delaunay, J. J. High-Q and tailorable Fano resonances in a one-dimensional metal-optical Tamm state structure: from a narrowband perfect absorber to a narrowband perfect reflector. *Adv Funct Mater* **31**, 2102183 (2021).

12.	Cong, S. *et al.* Noble metal-comparable SERS enhancement from semiconducting metal oxides by making oxygen vacancies. *Nat Commun* **6**, 7800 (2015).

13.	Wang, X. & Guo, L. SERS activity of semiconductors: crystalline and amorphous nanomaterials. *Angewandte Chemie* **132**, 4259–4267 (2020).

14.	Caldarola, M. *et al.* Non-plasmonic nanoantennas for surface enhanced spectroscopies with ultra-low heat conversion. *Nat Commun* **6**, 7915 (2015).

15.	Kuznetsov, A. I., Miroshnichenko, A. E., Brongersma, M. L., Kivshar, Y. S. & Luk'yanchuk, B. Optically resonant dielectric nanostructures. *Science* **354**, 2472 (2016).

16.	Deng, C. Z., Ho, Y. L., Clark, J. K., Yatsui, T. & Delaunay, J. J. Light switching with a metal-free chiral-sensitive metasurface at telecommunication wavelengths. *ACS Photonics* **7**, 2915–2922 (2020).

17.	Dagdeviren, O. E. *et al.* The effect of photoinduced surface oxygen vacancies on the charge carrier dynamics in TiO$_2$ Films. *Nano Lett* **21**, 8348–8354 (2021).

18. Wang, X. *et al.* Two-dimensional amorphous TiO$_2$ nanosheets enabling high-efficiency photoinduced charge transfer for excellent SERS activity. *J Am Chem Soc* **141**, 5856–5862 (2019).

19. Li, L. *et al.* Metal oxide nanoparticle mediated enhanced Raman scattering and its use in direct monitoring of interfacial chemical reactions. *Nano Lett* **12**, 4242–4246 (2012).

20. Qi, D., Lu, L., Wang, L. & Zhang, J. Improved SERS sensitivity on plasmon-free TiO$_2$ photonic microarray by enhancing light-matter coupling. *J Am Chem Soc* **136**, 9886–9889 (2014).

21. Alessandri, I. Enhancing Raman scattering without plasmons: unprecedented sensitivity achieved by TiO$_2$ shell-based resonators. *J Am Chem Soc* **135**, 5541–5544 (2013).

22. Wang, Z. *et al.* Self-Patterned CsPbBr$_3$ nanocrystal based plasmonic hot-carrier photodetector at telecommunications wavelengths. *Adv Opt Mater* **9**, 2101474 (2021).

23. Cambiasso, J., König, M., Cortés, E., Schlücker, S. & Maier, S. A. Surface-enhanced spectroscopies of a molecular Mmonolayer in an all-dielectric nanoantenna. *ACS Photonics* **5**, 1546–1557 (2018).

24. Spillane, S. M., Kippenberg, T. J. & Vahala, K. J. Ultralow-threshold Raman laser using a spherical dielectric microcavity. *Nature* **415**, 621–623 (2002).

25. Ohtaka, K. Enhanced Raman Scattering by two-dimensional array of polarizable spheres. *J Physical Soc Japan* **52**, 1457–1468 (1983).

26. Luther, J. M., Jain, P. K., Ewers, T. & Alivisatos, A. P. Localized surface plasmon resonances arising from free carriers in doped quantum dots. *Nat Mater* **10**, 361–366 (2011).

27. Romano, S. *et al.* Surface-enhanced Raman and fluorescence spectroscopy with an all-dielectric metasurface. *J Phys Chem C* **122**, 19738–19745 (2018).

28. Lagarkov, A. *et al.* SERS-active dielectric metamaterials based on periodic nanostructures. *Opt Express* **24**, 7133–7150 (2016).

29. Jahani, S. & Jacob, Z. All-dielectric metamaterials. *Nat Nanotechnol* **11**, 304 (2016).

30. Wang, P. *et al.* Molecular plasmonics with metamaterials. *Chem Rev* **122**, 2c00333 (2022).

31. Hüttenhofer, L. *et al.* Metasurface photoelectrodes for enhanced solar fuel generation. *Adv Energy Mater* **11**, 2102877 (2021).

32. Hsu, C. W., Zhen, B., Stone, A. D., Joannopoulos, J. D. & Soljačić, M. Bound states in the continuum. *Nat Rev Mater* **1**, 1–13 (2016).

33. Koshelev, K., Lepeshov, S., Liu, M., Bogdanov, A. & Kivshar, Y. Asymmetric Mmtasurfaces with high-Q resonances governed by bound states in the continuum. *Phys Rev Lett* **121**, 193903 (2018).

34. Kodigala, A. *et al.* Lasing action from photonic bound states in continuum. *Nature* **541**, 196–199 (2017).

35. Fernandez-Bravo, A. *et al.* Ultralow-threshold, continuous-wave upconverting lasing from subwavelength plasmons. *Nat Mater* **18**, 1172–1176 (2019).

36. Tittl, A., John-Herpin, A., Leitis, A., Arvelo, E. R. & Altug, H. Metasurface-based molecular biosensing aided by artificial intelligence. *Angewandte Chemie* **58**, 14810 (2019).

37. Tittl, A. *et al.* Imaging-based molecular barcoding with pixelated dielectric metasurfaces. *Science* **1109**, 1105–1109 (2018).

38. Carletti, L., Koshelev, K., Angelis, C. De & Kivshar, Y. Giant nonlinear response at the nanoscale driven by bound states in the continuum. *Phys Rev Lett* **121**, 33903 (2018).

39. Koshelev, K. *et al.* Subwavelength dielectric resonators for nonlinear nanophotonics. *Science* **367**, 288–292 (2020).

40. Christie, D., Lombardi, J. & Kretzschmar, I. Two-dimensional array of silica particles as a SERS substrate. *J Phys Chem C* **118**, 9114–9118 (2014).

41. Dyakov, S. A. *et al.* Photonic bound states in the continuum in Si structures with the self-assembled Ge nanoislands. *Laser Photon Rev* **15**, 2000242 (2021).

42. Kupriianov, A. S. *et al.* Metasurface engineering through bound states in the continuum. *Phys Rev Appl* **12**, 1–8 (2019).

43. Ra'Di, Y., Krasnok, A. & Alù, A. Virtual critical coupling. *ACS Photonics* **7**, 1468–1475 (2020).

44. Bell, S. E. J. *et al.* Towards reliable and quantitative surface-enhanced Raman scattering (SERS): from key parameters to good analytical practice. *Angewandte Chemie* **59**, 201908154 (2020).

45. Fan, Shanhui, Wonjoo Suh, and J. D. Joannopoulos. Temporal coupled-mode theory for the Fano resonance in optical resonators. *JOSA A* **20**, 569–572 (2003).

46. Maier, S. A. Plasmonic field enhancement and SERS in the effective mode volume picture. *Opt Express* **14**, 1957–1964 (2006).

47. Zhao, J., Lui, H., Mclean, D. I. & Zeng, H. Automated autofluorescence background subtraction algorithm for biomedical Raman spectroscopy. *Appl Spectrosc* **61**, 125–1232 (2007).

48. Rustichelli, D. *et al.* Validation of analytical methods in compliance with good manufacturing practice: A practical approach. *J Transl Med* **11**, 1–13 (2013).

49. Burtsev, V. *et al.* Detection of trace amounts of insoluble pharmaceuticals in water by extraction and SERS measurements in a microfluidic flow regime. *Analyst* **146**, 3686–3696 (2021).

50. Song, G., Gong, W., Cong, S. & Zhao, Z. Ultrathin two-dimensional nanostructures: surface defects for morphology-driven enhanced semiconductor SERS. *Angewandte Chemie* **60**, 5505–5511 (2021).

51. Wang, X., Shi, W., She, G. & Mu, L. Using Si and Ge nanostructures as substrates for surface-enhanced Raman scattering based on photoinduced charge transfer mechanism. *J Am Chem Soc* **133**, 16518–16523 (2011).

52. Lombardi, J. R. & Birke, R. L. Theory of surface-enhanced Raman scattering in semiconductors. *J Phys Chem C*. **118**, 11120–11130 (2014).

53. Hu, H. *et al.* Catalytic metasurfaces empowered by bound states in the continuum. *ACS Nano* **16**, 13057–13068 (2022).

54. Ding, S. Y., You, E. M., Tian, Z. Q. & Moskovits, M. Electromagnetic theories of surface-enhanced Raman spectroscopy. *Chem Soc Rev* **46**, 4042–4076 (2017).

55. Hasan, M. R. & Helleso, O. G. Metasurface supporting quasi-BIC for optical trapping and Raman-spectroscopy of biological nanoparticles. *Opt Express* **31**, 6782–6795 (2023).

56. Kristensen, P. T., Van Vlack, C. & Hughes, S. Generalized effective mode volume for leaky optical cavities. *Opt Lett* **37**, 1649–1651 (2012).

# Semiconductor Metasurfaces for Surface-enhanced Raman Scattering

*Haiyang Hu[1], Anil Kumar Pal[1], Alexander Berestennikov[1], Thomas Weber[1],*

*Andrei Stefancu[1], Emiliano Cortés[1], Stefan A. Maier[1,2,3], and Andreas Tittl[1\*].*

1. Chair in Hybrid Nanosystems, Nanoinstitute Munich, Faculty of Physics, Ludwig-Maximilians-Universität München, Königinstraße 10, 80539 München, Germany.

2. School of Physics and Astronomy, Monash University Clayton Campus, Melbourne, Victoria 3800, Australia.

3. The Blackett Laboratory, Department of Physics, Imperial College London, London SW7 2AZ, United Kingdom.

*E-mail: Andreas.Tittl@physik.uni-muenchen.de

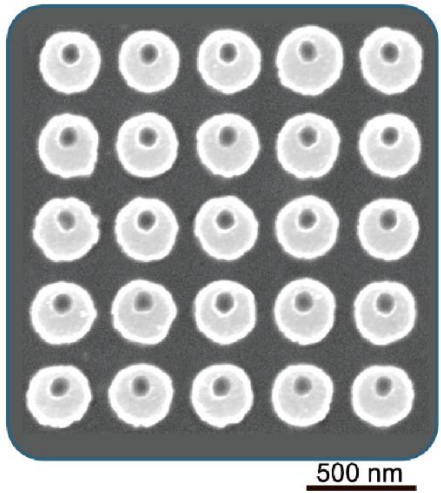 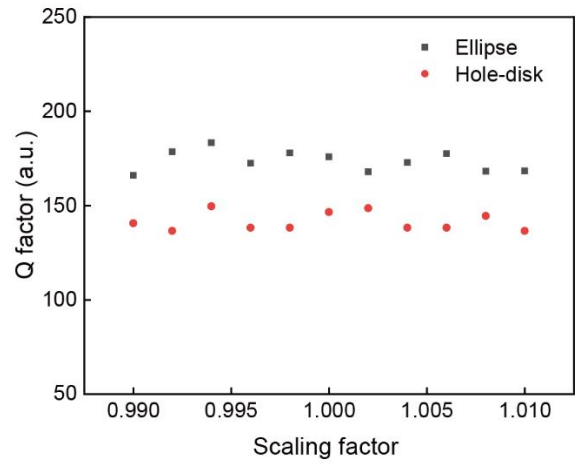

**Figure S1.** (a) SEM images of experimental hole-in-disk metasurfaces (scaling factor $S = 1$). (b) Quality factor (Q factor) of the quasi-BIC resonance with different geometries (two ellipses, and hole-disk) extracted from the experimental transmittance spectra with different scaling factor.

S

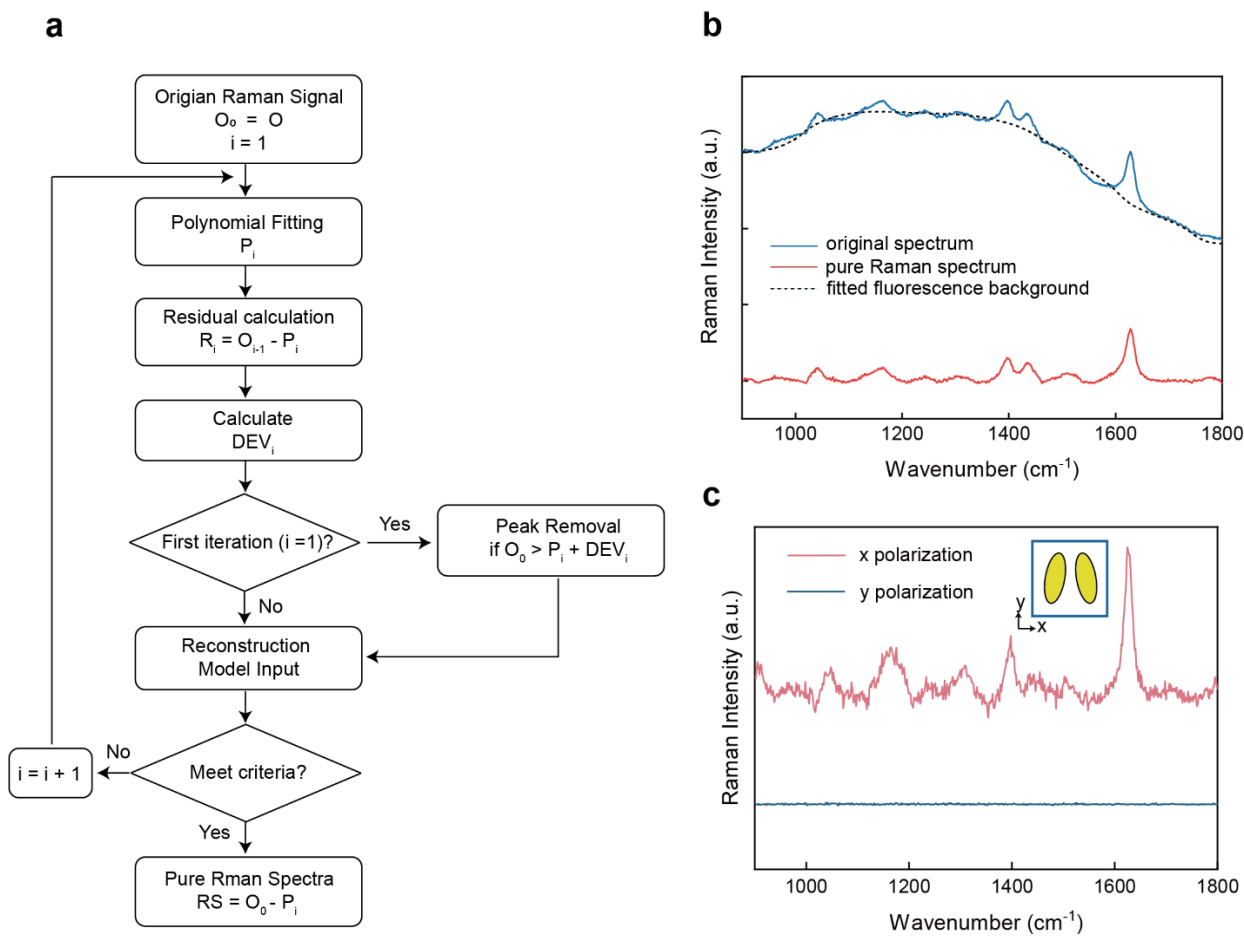

**Figure S2.** (a) Detailed diagram of the modified multi-polynomial fitting method. (b) The fluorescence background fitted by the modified multi-polynomial fitting method, and the extracted Raman spectrum. (c) Raman spectrums of the methylene blue (10E-7 M) were measured on the BIC metasurfaces with two ellipses geometry, which were excited by the x, and y polarized light respectively. The quasi-BIC resonance, which corresponds to an excitation laser wavelength of 633 nm, exhibits polarization dependency, specifically requiring the polarization of the incident light to be aligned along the x-axis for its activation for the SERS application.

## Derivation of TCMT equations

We start at the fundamental equations for the temporal evolution of a mode $a(t)$ in an open system [61], which can interact with the environment via the coupling constants $\boldsymbol{\kappa}$. The resonance frequency of the mode is given by $\omega_0$ and $\gamma = \gamma_{rad} + \gamma_{int}$ the sum of radiative and intrinsic loss rates.

$$\frac{da}{dt} = (i\omega_0 - \gamma)a + \boldsymbol{\kappa}^T \mathbf{s}_+ \quad (S1)$$

$$\mathbf{s}_- = C\mathbf{s}_+ + a\boldsymbol{\kappa} \quad (S2)$$

Our system is excited only through port 1 and allows for reflection and transmission, thus the explicit expressions for in- and outgoing waves ($\mathbf{s}_\pm$) are given by

$$\mathbf{s}_+ = \begin{pmatrix} s_{1+} \\ 0 \end{pmatrix}, \mathbf{s}_- = \begin{pmatrix} s_{1-} \\ s_{2-} \end{pmatrix}.$$

By acknowledging the conservation of energy, time-reversal symmetry and mirror symmetry with respect to the two ports, the port-coupling rates can be expressed as $\boldsymbol{\kappa} = \left(\sqrt{\gamma_{rad}}, \sqrt{\gamma_{rad}}\right)^T$.

The final parameter $C$ enables non-resonant energy transfer between ports 1 and 2 and is set here without loss of generality to

$$C = \begin{pmatrix} 0 & -1 \\ -1 & 0 \end{pmatrix}.$$

Assuming a harmonic time dependence ($a(t) \sim e^{i\omega t} \to d_t a = i\omega a$) equation (S1) can be solved for the mode amplitude

$$a = \frac{\sqrt{\gamma_{rad}}\, s_{1+}}{i(\omega - \omega_0) + \gamma_{int} + \gamma_{rad}}.$$

Because $a$ is normalized such that $|a|^2$ represents the energy inside the cavity [53] the electric field enhancement is proportional to the absolute value of the mode amplitude normalized to the incident field ($|E/E_0|^2 \sim |a/s_{1+}|^2$) and can be written on resonance ($\omega = \omega_0$) as

$$\left|\frac{E}{E_0}\right|^2 = b \cdot \frac{\gamma_{\text{rad}}}{(\gamma_{\text{int}} + \gamma_{\text{rad}})^2}, \quad (S3)$$

where $b$ is a proportionality constant used for later fits.

To derive an expression for the far-field absorbance $A$, we insert equation (S1) which is solved for $\mathbf{s}_+$ into (S2) and arrive at a general scattering-matrix formulation of our system

$$\mathbf{s}_- = S \mathbf{s}_+ \quad (S4)$$

$$S = \begin{pmatrix} s_{11} & s_{12} \\ s_{21} & s_{22} \end{pmatrix} := C + \frac{\boldsymbol{\kappa}\boldsymbol{\kappa}^{\text{T}}}{i(\omega - \omega_0) + \gamma_{\text{int}} + \gamma_{\text{rad}}}, \quad (S5)$$

where the reflection and transmission coefficients are defined as

$$r = \frac{s_{1-}}{s_{1+}} = s_{11}, \quad t = \frac{s_{2-}}{s_{1+}} = s_{21}$$

with explicit expressions given by

$$r = \frac{\gamma_{\text{rad}}}{i(\omega - \omega_0) + \gamma_{\text{int}} + \gamma_{\text{rad}}}, \quad t = -1 + \frac{\gamma_{\text{rad}}}{i(\omega - \omega_0) + \gamma_{\text{int}} + \gamma_{\text{rad}}}.$$

The absorbance is then retrieved via

$$A = 1 - R - T = 1 - |r|^2 - |t|^2 = \frac{2\gamma_{\text{int}}\gamma_{\text{rad}}}{(\omega - \omega_0)^2 + (\gamma_{\text{int}} + \gamma_{\text{rad}})^2} \quad (S6)$$

and yields at the critical coupling condition $\gamma_{\text{int}} = \gamma_{\text{rad}}$

$$A_{\max} = \frac{2\gamma^2}{4\gamma^2} = \frac{1}{2}.$$